\documentclass[preprint]{aastex} 

\begin{document}

\title{A HYBRID MODE MODEL OF THE BLAZHKO EFFECT, \\SHOWN TO ACCURATELY FIT \emph{KEPLER} DATA FOR RR LYR}

\author{Paul H. Bryant}
\affil{BioCircuits Institute (formerly Institute for Nonlinear Science), \\University of California, San Diego, La Jolla, CA 92093, USA}
\email{pbryant@ucsd.edu}

\begin{abstract}
The waveform for Blazhko stars can be substantially different during the ascending and descending parts of the Blazhko cycle.  A hybrid model, consisting of two component oscillators of the same frequency, is proposed as a means to fit the data over the entire cycle.  One component exhibits a sawtooth-like velocity waveform while the other is nearly sinusoidal.   One method of generating such a hybrid is presented: a nonlinear model is developed for the first overtone mode, which, if excited to large amplitude, is found to drop strongly in frequency and become highly non-sinusoidal.  If the frequency drops sufficiently to become equal to the fundamental frequency, the two can become phase locked and form the desired hybrid.   A relationship is assumed between the hybrid mode velocity and the observed light curve, which is approximated as a power series.  An accurate fit of the hybrid model is made to actual \emph{Kepler} data for RR Lyr.  The sinusoidal component may tend to stabilize the period of the hybrid which is found in real Blazhko data to be extremely stable.  It is proposed that the variations in amplitude and phase might result from a nonlinear interaction with a third mode, possibly a nonradial mode at 3/2 the fundamental frequency.  The hybrid model also applies to non-Blazhko RRab stars and provides an explanation for the light curve bump.  A method to estimate the surface gravity is also proposed.
\end{abstract}


\section{INTRODUCTION}
The Blazhko effect is a modulation of the pulsation amplitude that is seen in some RR Lyrae variable stars of RRab type.  Ever since \citet{Blazhko} discovered the effect, a wide variety of explanations have been put forth---recently reviewed by \citet{Kolenberg2012}.  Of particular relevance here are those which involve an interaction between two or more vibrational modes of the star, at least one of which is unstable \citep[see e.g.][]{Kollath, Buchler, Dziembowski, Van Hoolst, Moskalik}.

The waveform exhibited by Blazhko stars can look markedly different during the ascending and descending parts of the Blazhko cycle.  It is found that this behavior can be accurately reproduced with a hybrid model consisting of two component oscillators of identical frequency, one of which generates a sawtooth-like velocity waveform and the other is nearly sinusoidal.  The differing ascending and descending behavior is reproduced through variations in the amplitude and relative phase of the two components.

\begin{figure}
\begin{center}
\includegraphics[width=3.7in]{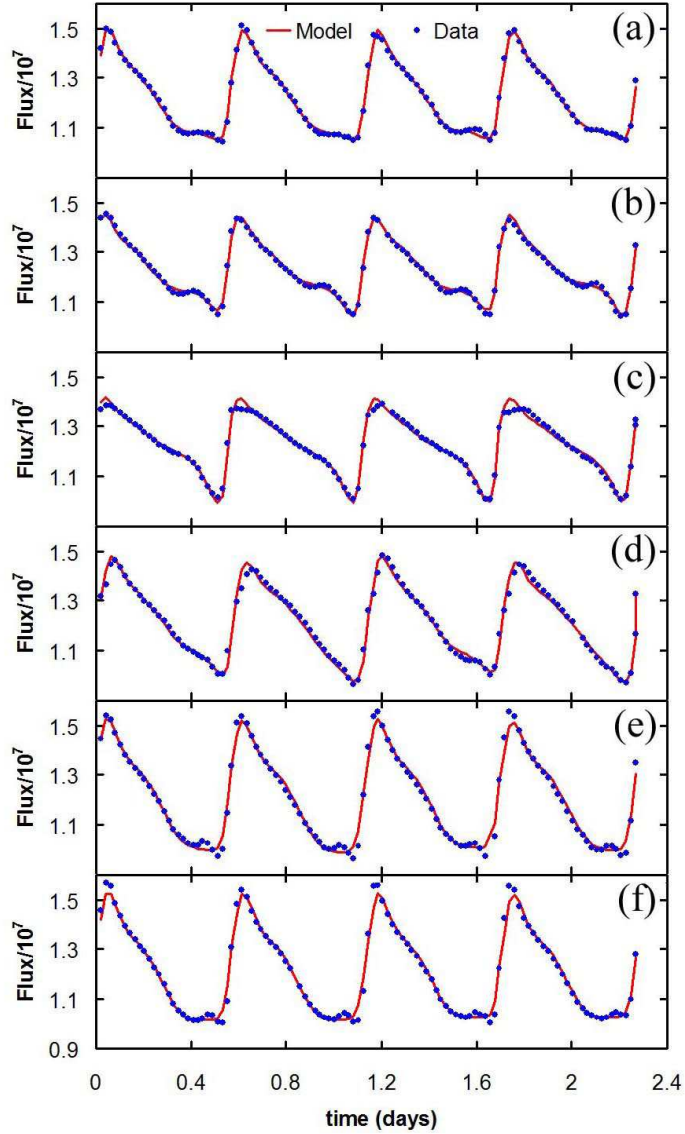}
\caption{\label{dvm}Comparison of \emph{Kepler} data (points) with the model (line), showing an accurate fit.  Data is for RR Lyr, collected in the second quarter of the \emph{Kepler} project.  Results are shown for four-cycle intervals at six different locations in the Blazhko cycle.  Relative to the beginning of the second quarter these have the following offsets:  (a) 12 cycles, (b) 24 cycles, (c) 36 cycles, (d) 48 cycles, (e) 60 cycles, and (f) 72 cycles.  This provides coverage of one Blazhko cycle.}
\end{center}
\end{figure}

A possible mechanism for a star to generate this hybrid mode is presented.  A model of the first-overtone mode is found to drop in frequency as the amplitude is increased.  If the overtone frequency drops sufficiently to reach the fundamental frequency, these modes can phase lock and produce the hybrid.  Using optimization techniques \citep[see e.g.][]{Press, Quinn}, the output of this hybrid model produces an accurate fit to \emph{Kepler} data for RR Lyr for the entire Blazhko cycle, as shown in Figure~\ref{dvm}.
Additional accuracy is achieved by adding small contributions from a mode with frequency 3/2 times the fundamental; its possible role in the Blazhko dynamics is discussed.  The rms error of the fit is typically about 3\% of the peak to peak amplitude.  These results are described in detail below.

Other possibilities may exist, either to generate the hybrid or to explain its ability to fit the data.  For example, a different pair of modes could be used, or perhaps some non-modal aspect of the dynamics could be shown to effectively generate the sinusoidal component.  Another possibility is a reversal of the roles where a strongly excited fundamental pulls down the frequency of the first overtone until they match.

The hybrid model will also fit data for non-Blazhko stars.  The waveform ``bump" is generated simply by combining the modes, similar to the resonance mechanism in bump Cepheids.  The bump moves and changes as is observed in the Blazhko cycle \citep{Guggenberger}.

In his shock-based Blazhko model, \citet{Gillet} also uses the first overtone, but with a ``transient" excitation rather than a continuous one.  \citet{Molnar} claims to observe the first overtone in RR Lyr spectra; however, this work does not consider that the modal frequencies may shift substantially (probably downward) due to the extremely disruptive effect of the main oscillation.  The hybrid components will \emph{not} appear as separate peaks since they have the same frequency.

The model for the overtone mode is equivalent to the one used by \citet{Stellingwerf1972} to study Cepheid variables.  It is a variant of the \citet{Baker} ``one-zone model."  \citet{Stellingwerf1986}  Figure 4 shows a case for which the period increased by about 60\%.  \citet{Stellingwerf1987} uses the same model to study the first overtone in RR Lyrae.  In a study using a solvable stellar model, \citet{Cox} finds a period increase of a factor of 1.54 for a case where the skewness of the velocity function appears roughly similar to that found in RRab stars---see his Figure 22(b), for $ w_+$ = 2, and his Table 4.

\begin{figure}
\begin{center}
\includegraphics[width=1.0\columnwidth]{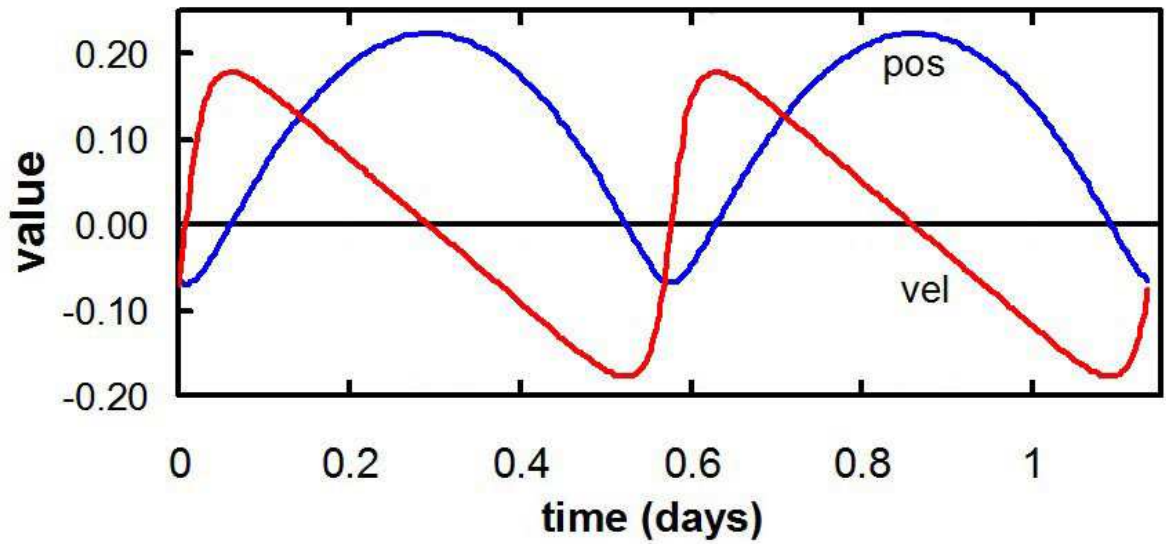}
\caption{\label{pv}Position $x(t)$ and velocity $v(t)$ for the overtone model.  Due to the high amplitude, the waveform is highly non-sinusoidal and the angular frequency has dropped from 15 to 11.1 rad day$^{-1}$, matching the frequency of the fundamental mode.  Model output is the same as used in Figure~\ref{dvm} part (a) and is nearly identical to that used in other parts.}
\end{center}
\end{figure}

The overtone model makes a gradual transition between two types of behavior:  at the low excitation limit it produces a sine wave and at the high excitation limit it produces a ``bouncing ball" wave.  In the latter case, the dynamics of the outer part of the star is primarily governed by gravity alone, except when it is near its most inward position.  At that point the outward pressure force is very strong, overcoming gravity and causing the motion to bounce.  Unlike the force in a simple harmonic oscillator that increases with displacement, gravity remains relatively constant with amplitude.  It is the weak nature of this restoring force that causes the period to increase.  Plotted against time this produces a succession of parabolic curves in a scalloped pattern.  The corresponding velocity function is a sawtooth-like wave that is seen in real data for RRab stars \citep[see e.g.][]{For} and certain other variable stars.  The ability to reproduce this waveform is a strong indication that the model is indeed capturing some of the essential physics of the real system.  The period of a bouncing ball increases with peak amplitude.  If the gravitational force can be approximated as constant throughout the cycle then in the high amplitude limit the period is proportional to the square root of the peak amplitude.  Typical waveforms for position and velocity from the model of the first overtone are shown in Figure~\ref{pv}, for the case where the frequency has dropped far enough to become equal to the frequency of the fundamental.

In the low amplitude limit, period increase can be examined in a \emph{universal} way that applies to all asymmetric oscillators.  Consider the equation for a simple harmonic oscillator and add the leading nonlinear correction to the restoring force, i.e., a term proportional to the square of the displacement.  Analysis of this equation shows that the fractional period increase is approximately fifteen times the square of the ratio of the second harmonic to the fundamental.  That ratio can be considered a measure of non-sinusoidality which is thus linked to period increase. 

It is known that the stellar oscillation period is extremely stable during the Blazhko cycle varying by less than 1\% \citep{Kolenberg2010,  Stellingwerf2013}.
This provides a strong argument for the hybrid model because the fundamental component remains sinusoidal (or very nearly so) and is thereby able to stabilize the frequency of the combination.  The overall amplitude changes are almost entirely due to changes in the fundamental's peak amplitude and relative phase.  The overtone remains essentially constant in amplitude and therefore does not shift in frequency during the Blazhko cycle.  If the hybrid was replaced by a single non-sinusoidal mode, the required amplitude changes would likely cause dramatic changes in the period of these oscillations which are not observed.

The relative phase is expected to be related to the rate of power transfer between the modes (via nonlinear mode coupling), and as will be shown below this appears to be consistent with amplitude changes of the fundamental providing another check on the validity of the hybrid model.

The sawtooth-like waveform of RR Lyrae and Cepheid variables appears to be a sign of gravity limited dynamics, and could therefore be used to estimate the surface gravity g.  The maximum inward slope could be obtained from radial velocity measurements \citep{For, Chadid} and used  as an estimate (or lower bound) for g.  Since g can also be determined from the radius and mass of a star, this could provide an additional check on the accuracy of those values.

\section{MODEL}
At low amplitudes, the vibrational dynamics of a star are known to reduce to a set of independent modes \citep[see e.g.][]{Unno},  each of which can be represented as a harmonic oscillator times a function of position that characterizes that mode.  For the first radial overtone there is a nodal surface at a certain radius.  The assumption is made that at high levels of excitation this spatial structure is maintained (at least approximately) and the main difference is that the forces governing the dynamics now become nonlinear and may be estimated by a simplified analysis of the system in question.

The primary forces acting on the dynamics of a star come from gravity and pressure.  Dissipation and excitation mechanisms will initially be ignored so that the dynamics will be conservative.  Results from fitting the model to \emph{Kepler} data show that the first overtone is the only mode that is driven strongly into the nonlinear regime, while the fundamental remains very nearly sinusoidal and is approximated here as a sine wave.  Although crude in its derivation, the model for the overtone mode is nevertheless good enough to produce dynamics that can achieve an accurate fit to real Blazhko data.  The model takes the form of an ordinary differential equation for a harmonic oscillator, but with the usual linear forcing term replaced with a nonlinear function $f(x)$ that, in a very simplified way, captures the essence of the gravity and pressure forces, i.e. $\dot x = v$ and $\dot v = f(x)$ where $x$ and $v$ are position and velocity.
A simple polytropic model \citep[see e.g.][]{Hansen} is used to estimate a typical pressure $P$, and it is assumed that a typical density is inversely proportional to a characteristic volume $V$:
\begin{equation}\label{polytrope}
P \propto 1/V^{(n+1)/n},
\end{equation}
where $n$ is the polytropic index.

Let $R_n$ be the nodal radius for the overtone mode.  The approach will be to focus on the dynamics above this radius.  It is assumed that there is an effective radius $r$ which has an equilibrium value $R_1$.  Let $x$ be the dynamical offset from equilibrium, so $r=R_1 + x$.  Approximate the volume of material above $R_n$ as proportional to $r^3 - R_n^3$.  Rescale so that $R_n = 1$.  Then the pressure force is
\begin{equation}\label{over_pforce}
f_p \propto r^2/ (r^3 - 1)^{(n+1)/n} = (R_1+x)^2/((R_1+x)^3-1)^{(n+1)/n}
\end{equation}
The gravitational force $f_g$ should go as $-1/r^2$, so the combined force $f$ will be of the form:
\begin{equation}\label{over_force}
f(x) = A(R_1+x)^2/((R_1+x)^3-1)^{(n+1)/n} - B/(R_1+x)^2,
\end{equation}
where $A$ and $B$ are constants.  Let $\omega$ be the small amplitude frequency for the overtone mode.  Require $f(0)=0$ and obtain $\omega$ from the gradient of the force at $x=0$, i.e., $\omega ^2 = -f^ \prime (0)$.  Solving for $A$ and $B$ it is found:
\begin{equation}\label{over_A}
A = \omega ^2 V_1 ^\gamma (3 \gamma R_1 ^4 V_1 ^{-1} -4R_1)^{-1}
\end{equation}
and
\begin{equation}\label{over_B}
B = \omega ^2 R_1 ^4 (3 \gamma R_1 ^4 V_1 ^{-1} -4R_1)^{-1},
\end{equation}
where $\gamma = (n+1)/n$ and $V_1 = R_1 ^3 -1$.
Typical results are shown in Figure~\ref{pv}.  The sharpness of the minima of the $x(t)$ plot is caused by the approach to the nodal surface.  For the case seen in the figure, the ``bounce zone" is relatively short in duration compared to the period, which is an indication that the dynamics will depend only slightly on the details of the pressure force.

\section{RESULTS AND DISCUSSION}
The approach is to attempt to fit actual data using a combination of modes, one of which is highly non-sinusoidal.  The data used is from RR Lyr, eponym of the class, from the \emph{Kepler} project (Kepler Input Catalog KIC 7198959, second quarter, long cadence corrected flux data) which covers about 150 stellar cycles and slightly over two Blazhko cycles.  The fitting has been done on intervals of four stellar cycles (111 \emph{Kepler} time-steps) and all usable data from the second quarter was included in the analysis.  It is found that all parts of the Blazhko cycle can be well fit with the combination of the fundamental and the first overtone, with the overtone being highly non-sinusoidal and having dropped in frequency to match the fundamental which is approximated as a sine wave.  The nodal surface of the overtone mode limits its available range of motion and causes it to become nonlinear at a lower level of energy than the fundamental.  The amplitude and relative phase of the fundamental vary gradually throughout the Blazhko cycle but are relatively constant throughout one four-cycle interval.  The first overtone varies little in amplitude throughout the Blazhko cycle, since any amplitude change would necessarily cause its frequency to move away from a match with the fundamental to which it is assumed to be phase locked by nonlinear interaction.

Additional accuracy is achieved by adding a small contribution from a third mode at 3/2 the fundamental frequency.  Evidence for such a mode is seen in the data \citep{Kolenberg2010}.  Other half-integer frequencies have been suggested for this mode \citep{Kollath, Buchler}, but 3/2 seemed to perform the best for fitting purposes.  If this mode is nonradial, it could help explain its small amplitude and seemingly erratic appearance in the light curve data.  A nonradial mode could also be a member of a large number of closely spaced modes.  If the evolution of the star caused it to sequentially be in resonance with consecutive members of this set, this could provide one explanation for the observed four-year meta-cycle \citep{Stellingwerf2013}.

Parameters are obtained by using an optimization algorithm \citep[see e.g.][]{Press, Quinn} to obtain the best mean square fit between the model and the data.  The parameters are held fixed for the entire data set, while the initial conditions (amplitude and phase of the modes) are re-optimized for the start of each four-cycle interval.  These changes reflect the effects of damping, excitation and energy transfer between the modes, which are assumed to be taking place on a slow timescale so that they can be ignored within a given four-cycle interval.  These effects are not directly included in the equations, though they likely will be in a more detailed future study.

Although a reasonably good fit can be obtained using the assumption that the light curve data is a constant plus a linear combination of the mode velocities, it is found that obtaining a good fit at a different interval requires some change in parameter values.  A solution to this problem was found by making the hypothesis that the light curve can be accurately reproduced by using a nonlinear function of the combined mode velocities.  This was achieved by using a power series expansion that included constant, linear, quadratic and cubic terms.  (No justification for this is currently known, other than to say that it seems to work remarkably well.)  These coefficients were also found using an optimization process, and are then fixed for the entire second quarter data set along with the other parameters.

\begin{figure}
\begin{center}
\includegraphics[width=1.0\columnwidth]{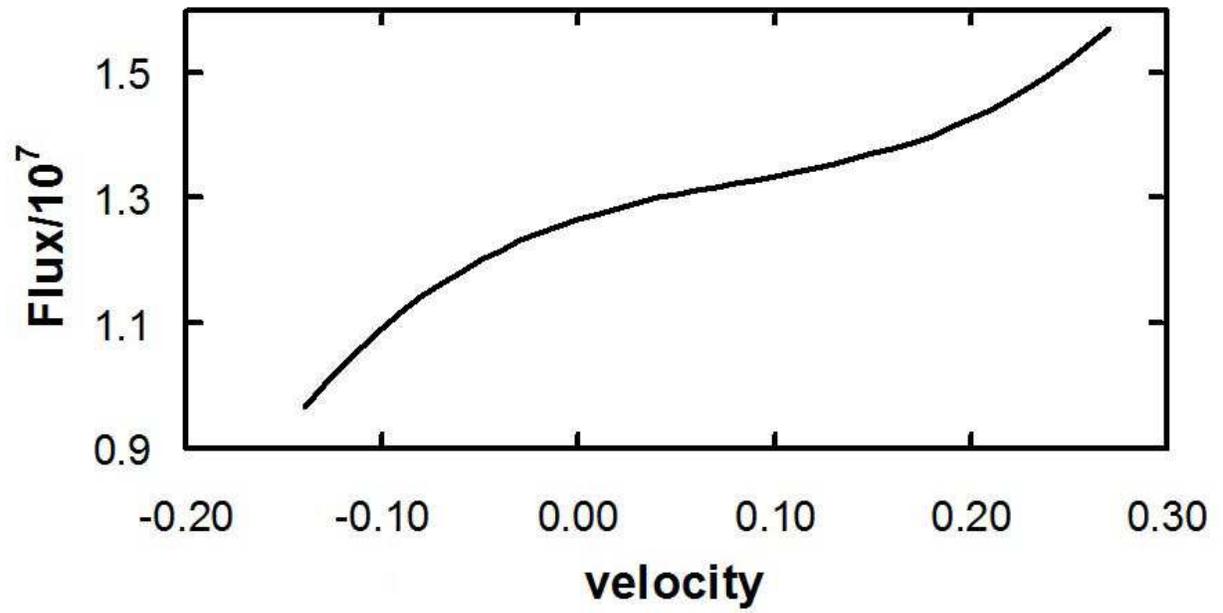}
\caption{\label{fluxfig}Flux function, calculated to optimize the fit of the model velocity output to the \emph{Kepler} flux data using a cubic expansion.}
\end{center}
\end{figure}

In Figture~\ref{dvm} is shown the fit of the model to the data at several different parts of the Blazhko cycle.  Note that these different intervals not only show changes in amplitude, but also changes in the shape of the waveform.  Generally the fit is excellent, although there are certain regions in which the data seems to have a slightly different motion than the model.  Attempting to improve on this is a goal for future study.  For modeling the data shown, some rescaling was done to keep the values closer to unity: the light flux was reduced by a factor of $10^7$, and time was measured in units of 0.1 days.  The polytropic index was set to 1.5.  The base angular frequency of the overtone mode was set to 1.5, and the frequency of the fundamental was set to 1.11 (radians per 0.1 day).  This corresponds to actual periods in days of 0.41888 and 0.56605 respectively, and base period ratio of 0.740, which is near the expected ratio for this type of star \citep[see e.g.][]{Soszynski}.  $R_1$ was set to 1.1, the output of the overtone model was scaled by the factor 1.03412, and the angular frequency of the third mode was optimized and ended up at 1.62444.  The fundamental mode velocity was approximated as pure sine wave and optimized in peak amplitude and initial phase for each interval independently.  The initial conditions for the overtone model (position and velocity) were also optimized for each interval independently.
\footnote{For Figure~\ref{dvm} parts (a)-(f) respectively, the intervals start at data points for which the truncated barycentric Julian dates are 55010.0596736, 55016.8642109, 55023.6687195, 55030.4731980, and 55037.2980786, 55044.0820593.  As found by optimization, the fundamental's peak amplitudes are 0.075003, 0.070437, 0.054499, 0.056218, 0.074069, and 0.078855; its initial phases are 1.70037, 2.23588, 2.50505, 1.53000, 1.44687, and 1.45994; the overtone's initial positions are $-0.066276$, $-0.067965$, $-0.068204$,  $-0.051784$, $-0.069204$, and $-0.067654$; its initial velocities are $-0.069905$, 0.047529, 0.040970,  $-0.139883$, $-0.009504$, and $-0.053195$, the third mode's peak amplitudes are 0.0008838, 0.0045300, 0.0008473, 0.0068498, 0.0022172, and 0.0009730; and its initial phases are 0.41714, $-0.67875$, 1.70241, 0.39568, 0.90341, and $1.22851$.}
The output of the overtone model was calculated by fourth order Runge-Kutta integration \citep[see, e.g.,][]{Press}, with a fixed half step that is one quarter of the \emph{Kepler} data time step (0.25 times 0.0204342 days).  The resulting $v(t)$ output is scaled by the factor given above, added to the fundamental velocity to produce a combined velocity $v_c(t)$.  (The small third mode correction was also added here as a sine wave.)  This is then processed by the flux function to produce the light output $L(t)$:
\begin{equation}\label{flux}
L(t)=k_0 + v_c(t) + k_2 v_c(t)^2 + k_3 v_c(t)^3
\end{equation}
where the coefficients optimized to the values: $k_0 = 1.267203$, $k_2 = -5.293310$ and $k_3 = 21.40176$.  This function is plotted for the range of interest in Figure~\ref{fluxfig}.

\begin{figure}
\begin{center}
\includegraphics[width=1.0\columnwidth]{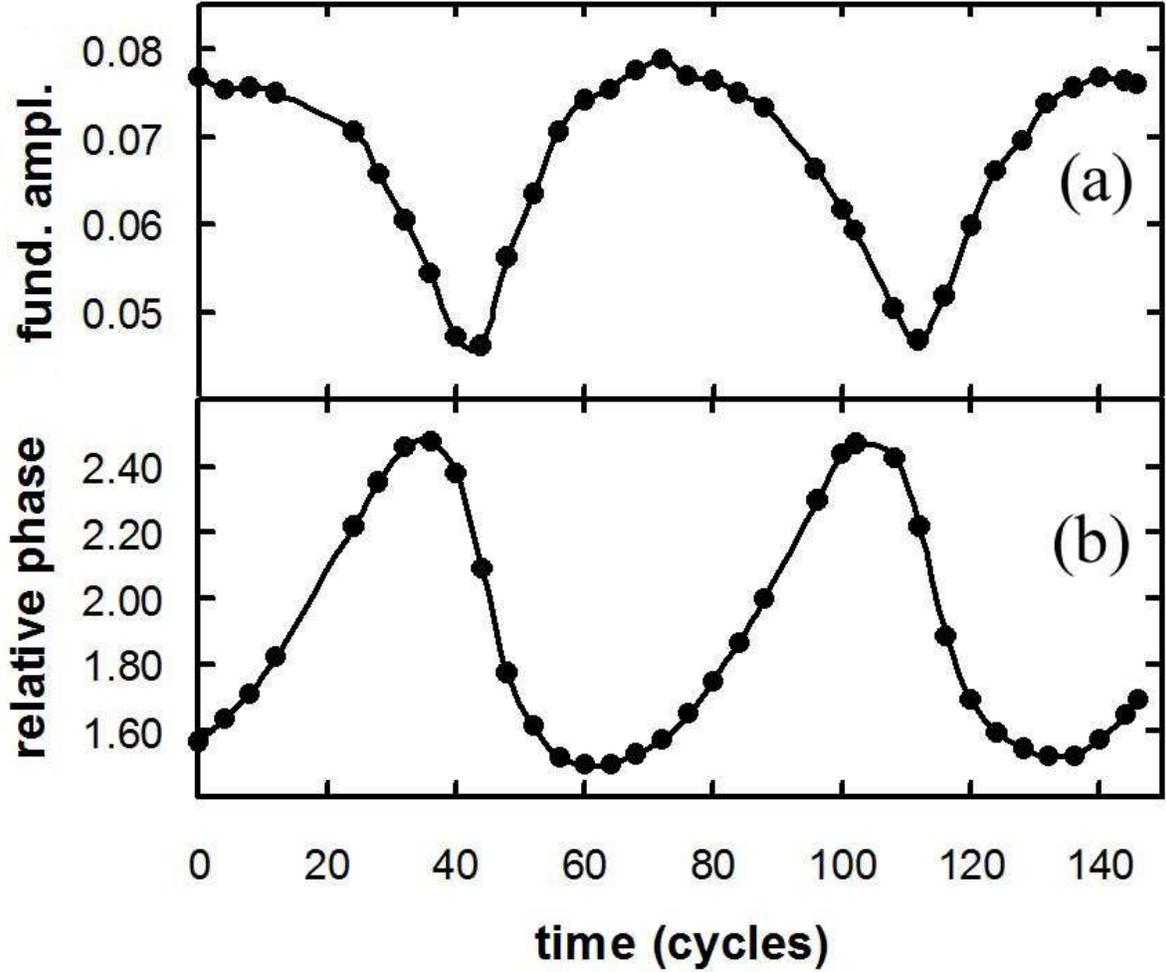}
\caption{\label{amp_phase}(a) Peak amplitude and (b) relative phase of the fundamental mode velocity as a function of time in stellar cycles since the start of the second quarter \emph{Kepler} data.  The fundamental velocity is approximated as a pure sine wave $v(t) = A \sin (\omega t + \phi)$, where $A$ is the amplitude.  $A$ and $\phi$ are determined by optimizing the fit of the model to the \emph{Kepler} data for each four-cycle interval.  The relative phase is found by taking the time at which the overtone velocity changes from negative to positive and subtracting the corresponding time for the fundamental and then multiplying by $2\pi$ divided by the period.  This is done at the crossing point nearest the center of each four-cycle interval.  Each dot in the figure is one calculation.}
\end{center}
\end{figure}

Figure~\ref{amp_phase}
shows the peak amplitude of the fundamental mode and its phase relative to the first overtone, as functions of time for the entire second quarter, obtained by the optimization process.  Both show cyclical behavior that corresponds to the Blazhko cycle.  Clearly the explanation for this behavior lies in the effects that have been omitted, i.e., the excitation, damping and nonlinear energy transfer between the fundamental and the overtone and possibly one or more additional modes.  Attempting to determine the details of these effects and interactions will be the subject of future work.

One hypothesis to explain the Blazhko effect is outlined as follows: the overtone is an ``unstable" mode, which acquires energy from the star and grows in amplitude until its frequency has dropped down to equal that of the fundamental to which it becomes phase locked.  The incoming energy acquired by the overtone still exceeds its energy loss due to damping, but instead of continuing to increase in amplitude, the excess energy of the overtone is transferred by nonlinear coupling to the fundamental and/or a third mode, possibly a nonradial mode at 3/2 the frequency.  The details of the coupling between these modes need further study, but in order to explain the Blazhko effect, there must be an inherent instability that causes this energy transfer to cycle back and forth between the fundamental and this third mode.  The energy transfer rate is related to the relative phase.  By comparing parts (a) and (b) of Figure~\ref{amp_phase} it is seen that the phase drops rapidly near cycle 43 and again near cycle 112 at the same times that the amplitude rapidly changes from decreasing to increasing.  Thus it would appear that this decrease in phase results in an increase in the energy transfer rate from the overtone to the fundamental causing the amplitude of the fundamental to increase.  This strong correlation between the amplitude and phase changes is further evidence that the hybrid mode concept is valid.

\acknowledgements

I thank Katrien Kolenberg for helpful discussion, and for introducing me to this interesting problem.  I thank Robert Stellingwerf for many helpful comments.

\end{document}